\documentclass[journal ]{new-aiaa}
\usepackage[utf8]{inputenc}
\usepackage{textcomp}
\usepackage{tikz}
\usepackage{graphicx}
\usepackage{amsmath}
\usepackage[version=4]{mhchem}
\usepackage{siunitx}
\usepackage{longtable,tabularx}
\title{Triton Aerogravity Assist Using a Flight-Proven, Inflatable Aerobrake for Neptune Orbit Capture}

\author{Jakob D. Brisby \footnote{Graduate Student, Department of Mechanical, Aerospace, and Biomedical Engineering, Email: jbrisby@vols.utk.edu} and James E. Lyne\footnote{Clinical Associate Professor, Department of Mechanical, Aerospace, and Biomedical Engineering, Email: jelyne@utk.edu}}
\affil{University of Tennessee at Knoxville, Knoxville, Tennessee, 37996}

\begin{document}

\maketitle
\begin{abstract}
    Previous work by our group has shown the potential for an aerogravity assist at Triton using a trailing ballute to capture a probe into orbit about Neptune. The current work extends that study by using the LOFTID aeroshell configuration, a flight-proven inflatable, to perform the aeromaneuver. Numerical simulations were carried out beginning at the atmospheric interface for a range of entry velocities (3.0-14.0 km/s) and angles (45-56 degrees). The spin-stabilized vehicle was assumed to fly at zero angle of attack, producing no lift, and is capable of reaching outbound conditions sufficient to capture into orbit about Neptune without violating the vehicle's aerothermal limits or penetrating too deeply into Triton’s atmosphere. 
\end{abstract}

\section{Introduction}
\lettrine{T}{he} idea of using an aerogravity assist (AGA) at Titan for orbital capture about Saturn was proposed at least as early as 1990 by McRonald and Randolph \cite{1990aiaa.meetW....M}. Numerous studies since that time have explored the concept more fully with regard to guidance and control, aerothermal heating, and other issues \cite{randolph-1992,ramsey-2003,ramsey-2004,ramsey-2006,ramsey-2008,spilker-2009,2021BAAS...53d.382A,2021BAAS...53d.489T,lu-2020,spilker-2019}.  Because of its high atmospheric density coupled with a large-scale height and weak gravitational field, Titan is an ideal location to conduct such a maneuver, and studies have indicated that low L/D, capsule-type configurations would likely be sufficient.

In 2004, our group proposed using the same method at Triton to capture a spacecraft into orbit about Neptune \cite{ramsey-2004,ramsey-2005}. Triton is of specific interest for such an AGA capture due to its retrograde orbit about Neptune; this means an aerocapture using Neptune's atmosphere into an orbit amenable to subsequent manipulation via repeated Triton flybys requires an atmospheric entry opposite to the direction of Neptune's rotation, resulting in higher relative velocities and a more severe aerothermal environment \cite{lu-2020}. Capturing into Neptune's orbit via a Triton AGA however, would reduce the atmospheric entry speed, since the spacecraft would not enter as deeply into Neptune's gravity well and would be traveling roughly in the same direction as Triton's motion. Triton, however, has an atmospheric density four orders of magnitude lower than Titan, making a successful AGA much more challenging than it would be in the Saturn system. Largely as a result of this low atmospheric density, our previous work indicated that a capsule would probably not be sufficient, but a ballute might carry out the maneuver successfully. In addition, it was suggested that “low ballistic coefficient aeroshells, such as those with inflatable skirts” should be considered in future studies \cite{ramsey-2005}.

\section{Background and Objectives}

The technology required for such a low ballistic coefficient vehicle has recently been brought to maturity with NASA's successful flight test of the LOFTID aeroshell, a 6-m diameter, 70-degree half-angle, spin-stabilized, inflatable sphere-cone which shares a forebody shape with the heritage Viking entry vehicle. LOFTID entered Earth's atmosphere at almost 8 km/s in November of 2022 on an uncontrolled trajectory after being launched from Vandenberg Space Force Base aboard an Atlas V rocket \cite{amert-2020,cheatwood-2021}.  The vehicle had a maximum allowable heating rate of approximately 60 W/cm$^{2}$. 

In the present work, we seek to apply the technology developed for LOFTID to a Triton AGA used for orbital capture about Neptune. To minimize potential development costs, the configuration considered is essentially identical to the flight test vehicle in mass (1100 kg), size (6 m diameter, 28.27 m$^{2}$ reference area), and aerodynamics (drag coefficient 1.69 at angle of attack of zero).  For this preliminary study, it is assumed that the axisymmetric vehicle flies at an angle of attack of zero and hence produces no lift, as was the case with the LOFTID entry.

Triton's atmosphere is primarily N$_2$ at 99.97\%, methane at 0.0036\%, and other trace species\cite{krasnopolsky-1993}. Stellar and solar occultation studies \cite{krasnopolsky-1993,elliot-2000}, have been used to determine atmospheric density as a function of altitude as shown in Fig. 1, indicating a density at the surface on the order of 10$^{-4}$ kg/m$^3$. Based on the atmospheric number density and the size of the vehicle, continuum aerodynamics should be applicable below approximately 70 km altitude.

\begin{figure}[hbt!]
\centering
\includegraphics[width=0.8\textwidth]{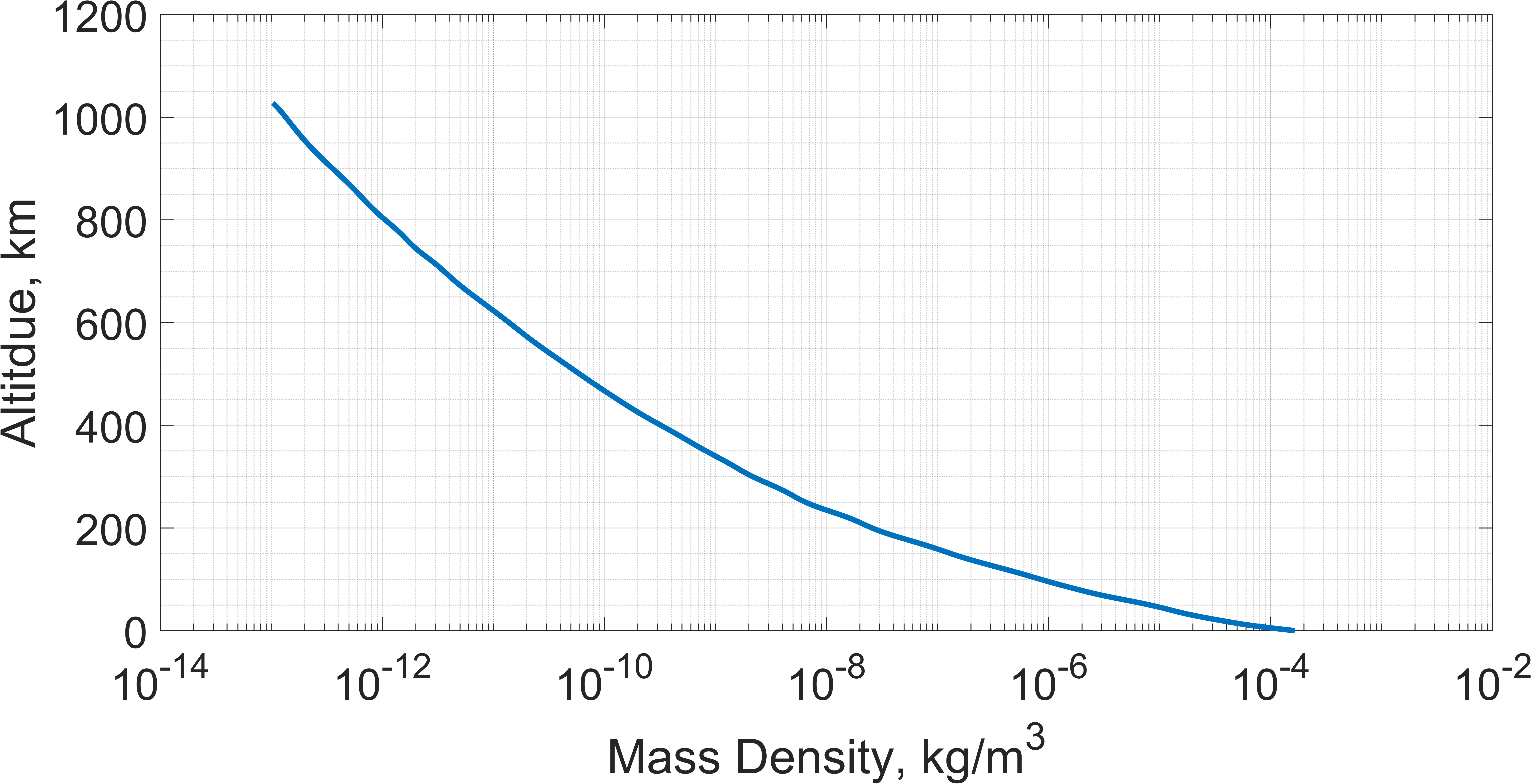}
\caption{Mass density profile through Triton's atmosphere from Strobel \cite{strobel-2017}.} \label{fig:strob3}
\end{figure}

\section{Methodology}
Triton atmospheric trajectories were modeled using POST (The Program to Optimize Simulated Trajectories) \cite{POST}. Simulations were begun and entry conditions specified at an altitude of 1029 km, assuming no rotation of Triton and a due east, equatorial entry. The atmospheric entry speed at Triton depends on the probe's arrival excess speed at the Neptune system as well as the relative alignment of the probe's velocity vector and Triton's orbital motion at the time of the intercept. A diagram of the relative velocity vectors is shown in Fig. 2 where V$_{P,i}$ is the velocity of the probe in Neptune's reference frame when it's at Triton's orbital radius before the AGA, V$_{Tr}$ is Triton’s orbital velocity, V$_{P,f}$ is velocity of the probe with respect to Netpune after AGA, $\delta$ is the turn angle of V$_{\infty}$ during AGA, and $\phi$ is the angle between Triton's velocity vector and outbound V$_{\infty,o}$. Based on prior studies of Neptune aerocapture in conjunction with reasonable limits on the anticipated aerothermal heating, we considered Triton atmospheric entry speeds ranging from 3.0 to 14.0 km/s and atmospheric entry angles between -45.0 and -56.0 degrees.

\begin{figure}[hbt!]
\centering
\tikzset{every picture/.style={line width=0.75pt}} 

\begin{tikzpicture}[x=0.5pt,y=0.5pt,yscale=-1,xscale=1]

\draw [color={rgb, 255:red, 0; green, 0; blue, 0 }  ,draw opacity=1 ][line width=1.5]    (180,297.5) -- (547.85,297.25) ;
\draw [shift={(177,297.5)}, rotate = 359.96] [color={rgb, 255:red, 0; green, 0; blue, 0 }  ,draw opacity=1 ][line width=1.5]    (14.21,-4.28) .. controls (9.04,-1.82) and (4.3,-0.39) .. (0,0) .. controls (4.3,0.39) and (9.04,1.82) .. (14.21,4.28)   ;
\draw [color={rgb, 255:red, 0; green, 0; blue, 0 }  ,draw opacity=1 ][line width=1.5]    (432.47,91.12) -- (547.85,297.25) ;
\draw [shift={(431,88.5)}, rotate = 60.76] [color={rgb, 255:red, 0; green, 0; blue, 0 }  ,draw opacity=1 ][line width=1.5]    (14.21,-4.28) .. controls (9.04,-1.82) and (4.3,-0.39) .. (0,0) .. controls (4.3,0.39) and (9.04,1.82) .. (14.21,4.28)   ;
\draw [color={rgb, 255:red, 0; green, 0; blue, 0 }  ,draw opacity=1 ][line width=1.5]    (179.32,295.59) -- (431,88.5) ;
\draw [shift={(177,297.5)}, rotate = 320.55] [color={rgb, 255:red, 0; green, 0; blue, 0 }  ,draw opacity=1 ][line width=1.5]    (14.21,-4.28) .. controls (9.04,-1.82) and (4.3,-0.39) .. (0,0) .. controls (4.3,0.39) and (9.04,1.82) .. (14.21,4.28)   ;
\draw [color={rgb, 255:red, 0; green, 0; blue, 0 }  ,draw opacity=1 ][line width=1.5]  [dash pattern={on 1.69pt off 2.76pt}]  (360.18,254.74) -- (431,88.5) ;
\draw [shift={(359,257.5)}, rotate = 293.08] [color={rgb, 255:red, 0; green, 0; blue, 0 }  ,draw opacity=1 ][line width=1.5]    (14.21,-4.28) .. controls (9.04,-1.82) and (4.3,-0.39) .. (0,0) .. controls (4.3,0.39) and (9.04,1.82) .. (14.21,4.28)   ;
\draw [color={rgb, 255:red, 0; green, 0; blue, 0 }  ,draw opacity=1 ][line width=1.5]  [dash pattern={on 1.69pt off 2.76pt}]  (361.94,258.12) -- (547.85,297.25) ;
\draw [shift={(359,257.5)}, rotate = 11.89] [color={rgb, 255:red, 0; green, 0; blue, 0 }  ,draw opacity=1 ][line width=1.5]    (14.21,-4.28) .. controls (9.04,-1.82) and (4.3,-0.39) .. (0,0) .. controls (4.3,0.39) and (9.04,1.82) .. (14.21,4.28)   ;
\draw  [draw opacity=0][line width=1.5]  (526.22,248.53) .. controls (528.06,246.15) and (530.56,244.02) .. (533.7,242.25) .. controls (541.19,238.04) and (550.86,236.74) .. (560.31,238.05) -- (558,263) -- cycle ; \draw [line width=1.5]    (528.22,246.29) .. controls (529.75,244.8) and (531.58,243.44) .. (533.7,242.25) .. controls (541.19,238.04) and (550.86,236.74) .. (560.31,238.05) ;  \draw [shift={(526.22,248.53)}, rotate = 322.9] [color={rgb, 255:red, 0; green, 0; blue, 0 }  ][line width=1.5]    (14.21,-4.28) .. controls (9.04,-1.82) and (4.3,-0.39) .. (0,0) .. controls (4.3,0.39) and (9.04,1.82) .. (14.21,4.28)   ;
\draw  [draw opacity=0][line width=1.5]  (512.97,333.5) .. controls (508.44,330.66) and (504.91,325.83) .. (503.16,319.31) .. controls (501.62,313.59) and (501.66,307.35) .. (502.99,301.29) -- (528.89,303.36) -- cycle ; \draw [line width=1.5]    (512.97,333.5) .. controls (508.44,330.66) and (504.91,325.83) .. (503.16,319.31) .. controls (501.87,314.51) and (501.69,309.33) .. (502.45,304.21) ; \draw [shift={(502.99,301.29)}, rotate = 93.93] [color={rgb, 255:red, 0; green, 0; blue, 0 }  ][line width=1.5]    (14.21,-4.28) .. controls (9.04,-1.82) and (4.3,-0.39) .. (0,0) .. controls (4.3,0.39) and (9.04,1.82) .. (14.21,4.28)   ; 
\draw  [draw opacity=0][line width=1.5]  (453.69,138.02) .. controls (451.34,142.71) and (446.81,146.46) .. (440.33,148.45) .. controls (429.42,151.8) and (415.83,149.31) .. (404.94,142.84) -- (421,123) -- cycle ; \draw [line width=1.5]    (452.12,140.59) .. controls (449.54,144.07) and (445.57,146.83) .. (440.33,148.45) .. controls (430.24,151.55) and (417.85,149.65) .. (407.43,144.23) ; \draw [shift={(404.94,142.84)}, rotate = 23.26] [color={rgb, 255:red, 0; green, 0; blue, 0 }  ][line width=1.5]    (14.21,-4.28) .. controls (9.04,-1.82) and (4.3,-0.39) .. (0,0) .. controls (4.3,0.39) and (9.04,1.82) .. (14.21,4.28)   ; \draw [shift={(453.69,138.02)}, rotate = 135.65] [color={rgb, 255:red, 0; green, 0; blue, 0 }  ][line width=1.5]    (14.21,-4.28) .. controls (9.04,-1.82) and (4.3,-0.39) .. (0,0) .. controls (4.3,0.39) and (9.04,1.82) .. (14.21,4.28)   ;
\draw  [draw opacity=0][line width=1.5]  (388,182.97) .. controls (382.28,185.37) and (375.72,185.96) .. (369.17,184.22) .. controls (358.14,181.3) and (350,172.55) .. (347.23,162.12) -- (376,155) -- cycle ; \draw [line width=1.5]    (385.1,184.01) .. controls (380.11,185.5) and (374.64,185.67) .. (369.17,184.22) .. controls (359.13,181.57) and (351.48,174.08) .. (348.1,164.89) ; \draw [shift={(347.23,162.12)}, rotate = 64.52] [color={rgb, 255:red, 0; green, 0; blue, 0 }  ][line width=1.5]    (14.21,-4.28) .. controls (9.04,-1.82) and (4.3,-0.39) .. (0,0) .. controls (4.3,0.39) and (9.04,1.82) .. (14.21,4.28)   ; \draw [shift={(388,182.97)}, rotate = 169.26] [color={rgb, 255:red, 0; green, 0; blue, 0 }  ][line width=1.5]    (14.21,-4.28) .. controls (9.04,-1.82) and (4.3,-0.39) .. (0,0) .. controls (4.3,0.39) and (9.04,1.82) .. (14.21,4.28)   ;

\draw (260,163.4) node [anchor=north west][inner sep=0.75pt]  [font=\Large]  {$V_{\infty ,i}$};
\draw (550,239.4) node [anchor=north west][inner sep=0.75pt]  [font=\Large]  {$\theta _{i}$};
\draw (424,117.4) node [anchor=north west][inner sep=0.75pt]  [font=\Large]  {$\phi $};
\draw (366,145.4) node [anchor=north west][inner sep=0.75pt]  [font=\Large]  {$\delta $};
\draw (357,300.4) node [anchor=north west][inner sep=0.75pt]  [font=\Large]  {$V_{P,i}$};
\draw (439,245.4) node [anchor=north west][inner sep=0.75pt]  [font=\Large]  {$V_{P,f}$};
\draw (389,194.4) node [anchor=north west][inner sep=0.75pt]  [font=\Large]  {$V_{\infty ,o}$};
\draw (494,165.4) node [anchor=north west][inner sep=0.75pt]  [font=\large]  {$V_{Tr} =4.39$};

\end{tikzpicture}
\caption{Diagram of relative velocity vectors during the Triton approach.} \label{fig:velVect}
\end{figure}
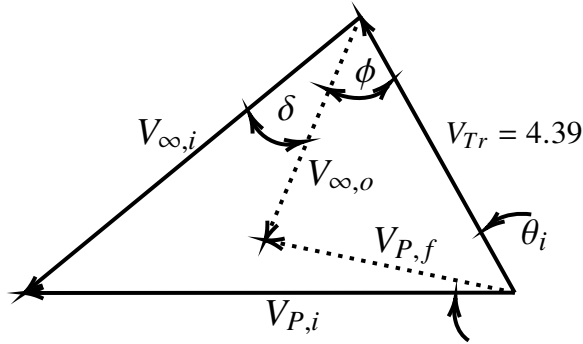

Convective and radiative heat transfer rates were calculated for the trajectories based on the methods described by Tauber \cite{1989STIA...9021585T}, Tauber and Sutton \cite{1991JSpRo..28...40T}, and Sutton and Hartung \cite{sutton-1990} for flight in Earth’s atmosphere. Since Triton’s atmosphere is primarily N$_{2}$, these methods should provide a reasonable first pass estimate.

\section{Results}
\begin{figure}[hbt!]
\centering
\includegraphics[width=0.6\textwidth]{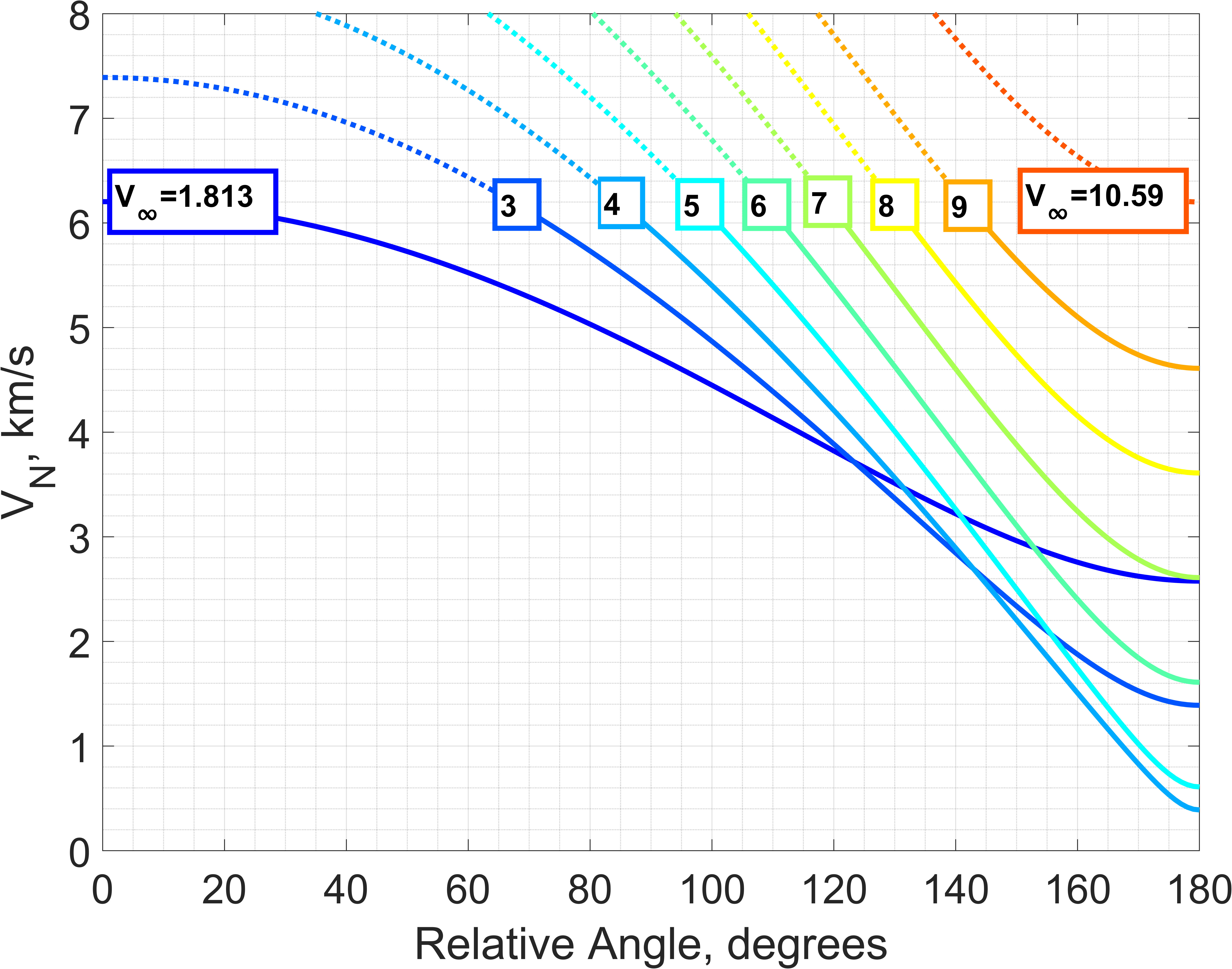}
\caption{Neptune centered velocity magnitude as a function of alignment of V$_{\infty,o}$ with Triton's velocity vector.} \label{fig:captVel}
\end{figure}

Since Triton's orbit about Neptune is nearly circular at a velocity of 4.39 km/s, any vehicle leaving Triton with an outbound excess speed of less than 1.813 km/s will be captured into the Neptune system regardless of its direction of travel, while any vehicle leaving Triton with an outbound excess speed over 10.59 km/s will not be captured into the Neptune system. As shown in Fig. \ref{fig:captVel}, values of outbound excess speed well above 1.813 km/s have potential for closed orbits about Neptune, depending on the alignment of the outbound velocity vector with Triton's orbital motion. Solid lines represent when conditions allow for Neptunian capture orbit without adjusting the velocity vector. The dotted lines represent outbound conditions from Triton that will escape the Neptunian system unless another $\Delta$V maneuver is employed.

\begin{figure}[hbt!]
\centering
\includegraphics[width=0.6\textwidth]{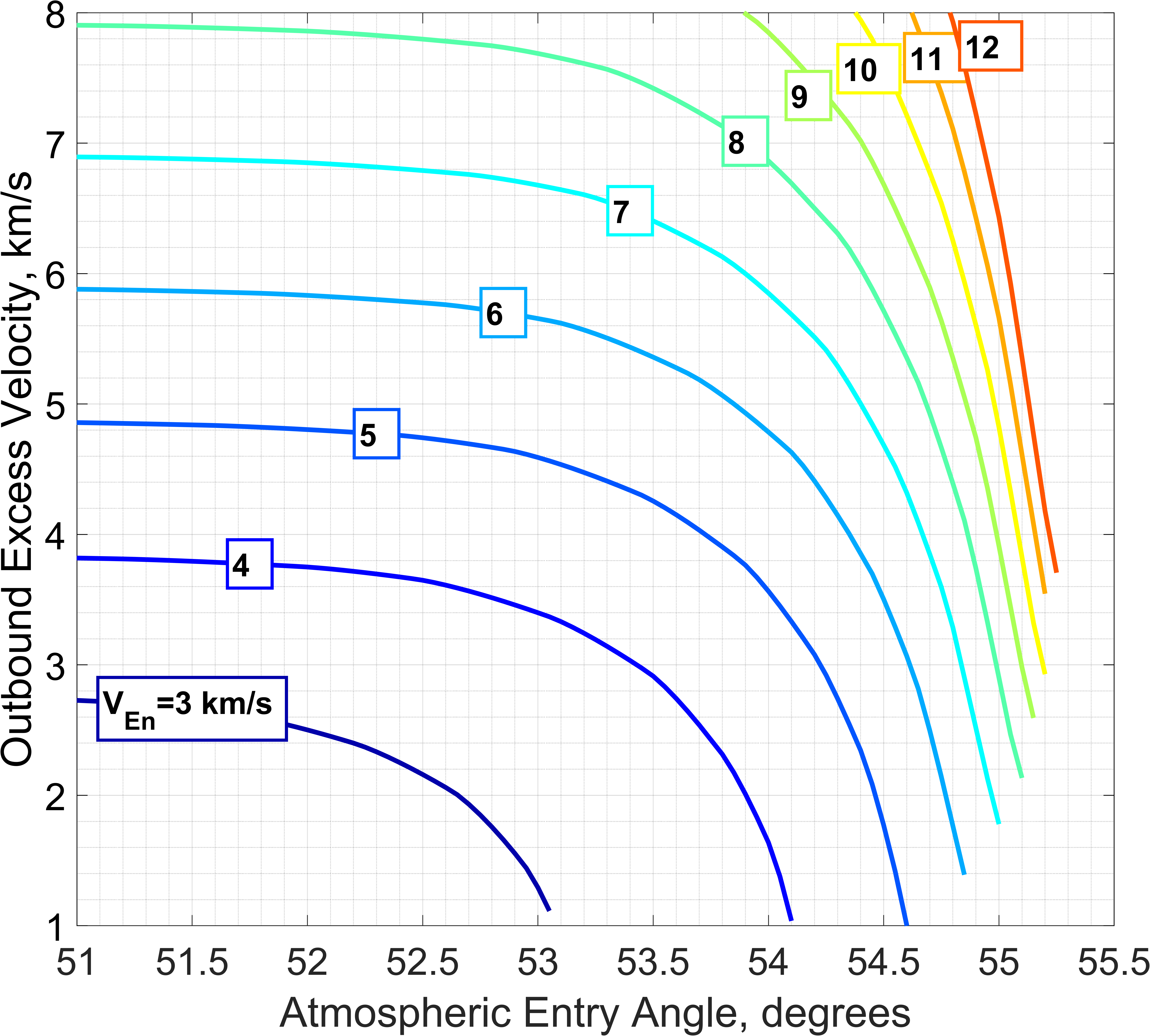}
\caption{Triton outbound excess velocity as a function of entry conditions.} \label{fig:vinfOut_entry}
\end{figure}

\begin{figure}[hbt!]
\centering
\includegraphics[width=0.6\textwidth]{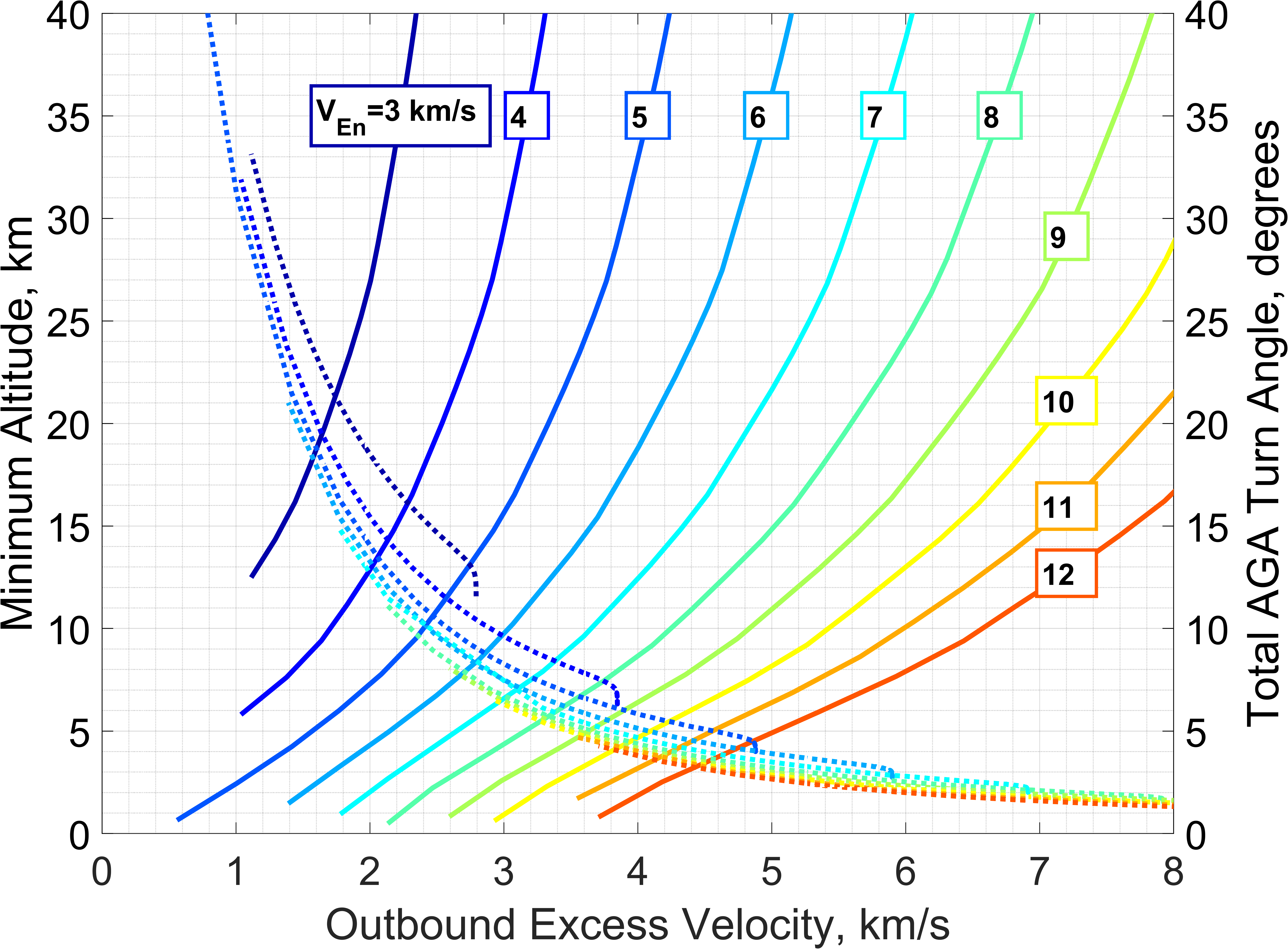}
\caption{Minimum altitude (solid lines) and turn angle (dashed lines) as a function of outbound excess speed and atmospheric entry speed.} \label{fig:turn_altVinf}
\end{figure}

\begin{figure}[hbt!]
\centering
\includegraphics[width=0.6\textwidth]{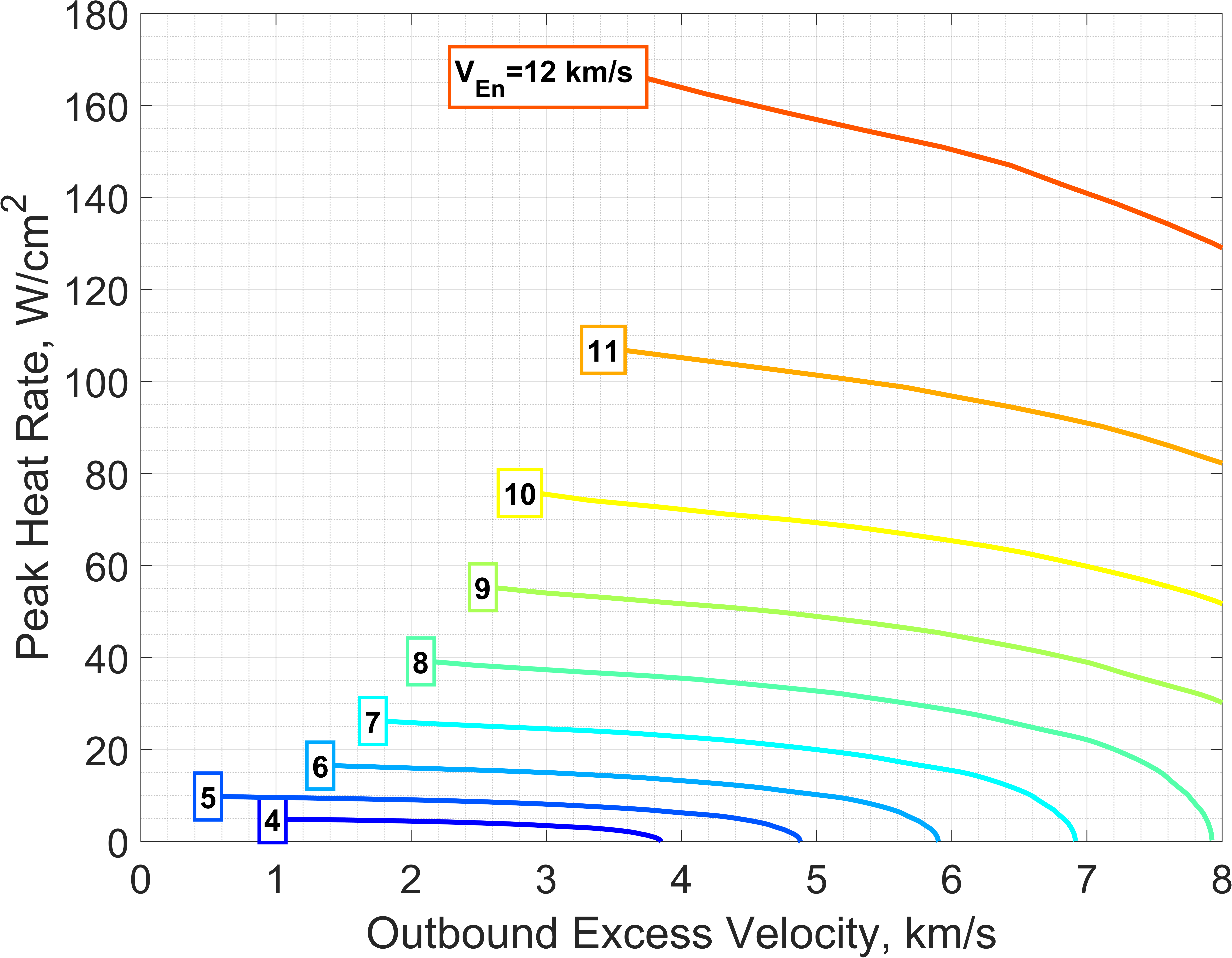}
\caption{Peak stagnation point heating rate as a function of outbound excess speed and atmospheric entry velocity.} \label{fig:pkqdot}
\end{figure}

Figure \ref{fig:vinfOut_entry} shows the outbound excess speed (V$_{\infty,o}$) of the LOFTID vehicle as a function of the Triton atmospheric entry velocity and entry angle, while Fig. \ref{fig:turn_altVinf} shows the minimum altitude and total AGA turn angle as a function of V$_{\infty,o}$ for the same range of entry speeds. The outbound excess speed and turn angle are very sensitive to the atmospheric entry angle, and this becomes more pronounced at higher entry velocities. Figure \ref{fig:pkqdot} shows the peak stagnation point heating rate, assuming a nose radius of 1.5m, as a function of entry velocity and V$_{\infty,o}$. As would be expected, greater energy loss during the AGA maneuver requires a deeper penetration into the atmosphere, and this results in a higher peak heating rate and a larger turn angle.

Careful examination of these results reveals cases in which the LOFTID configuration appears to be capable of providing an adequate energy loss during a Triton AGA maneuver to capture the vehicle into Neptune's orbit without violating the vehicle's aerothermal limits or reasonable altitude constraints. For example, an 18.84-year direct Earth to Neptune trajectory arrives at Neptune with a hyperbolic excess speed of 5.44 km/s. If Triton intercept occurs with an angle between the arrival velocity vector and Triton's orbital motion vector of 0 to 60 degrees, Triton atmospheric entry speeds would range between 4.01 and 7.23 km/s. Assuming the nominal atmospheric density profile and a minimum allowable altitude of 10 km during the AGA, outbound excess speeds range from 1.7 to 3.79 km/s, and turn angles from 18.5 to 5 degrees,  resulting in capture into a closed orbit about Neptune. Faster direct transits will have a narrower range of intercept angles ($\theta_{i}$) for which a successful capture can be accomplished, while slower transits will have a wider range of potentially successful intercept angles.

\section{Conclusion}
For certain arrival scenarios, the LOFTID configuration appears to be capable of providing an adequate energy loss during a Triton AGA maneuver to capture the vehicle into a Neptune orbit. Such a maneuver would likely require a transit time to Neptune of 15 years or more and a low peripase altitude during the AGA.  However, given the lack of prominent variation in surface elevation on Triton, such a low altitude maneuver may be feasible.

Since in this study the axisymmetric vehicle is assumed to fly at zero angle of attack, it has a lift coefficient of zero, and, as a result, there is a one-to-one relationship between the atmospheric entry angle and both the energy loss during the AGA and the outbound excess speed. This lack of control authority makes the final orbit achieved very sensitive to dispersions in the entry angle or in the atmospheric density profile encountered. Moreover, unlike a ballute, the current LOFTID configuration is not designed to detach. Future studies are needed to examine the entry corridor width; the necessary control authority could be provided either by flying the vehicle at a non-zero angle of attack in a lifting configuration as was done by the IRVE-3 vehicle \cite{olds-2013} or by making modifications that would allow a large inflatable skirt to separate from a much smaller central capsule. The effects of off nominal atmospheric density profiles must also be examined.

\section*{Acknowledgments}
The authors would like to thank Dr. Neil Cheatwood for providing information about the LOFTID flight. 

\bibliography{refs}

\end{document}